\documentclass[a4paper,11pt]{article}
\usepackage{pos}
\usepackage{bbold}

\newcommand{\ket}[1]{\ensuremath{| {#1} \rangle }}
\newcommand{\bra}[1]{\ensuremath{\langle {#1} |}}

\renewcommand{\d}{\mathrm{d}}

\title{Smeared $R$-ratio in isospin symmetric QCD with Low Mode Averaging}

\author[a]{Simone Bacchio}
\author[b]{Alessandro De~Santis} 
\author[c]{Antonio Evangelista}
\author[c]{Roberto Frezzotti} 
\author[d]{Giuseppe Gagliardi}
\author[e]{Marco Garofalo}
\author*[c]{Francesca Margari} 
\author[a]{Ferenc Pittler}
\author[f]{Francesco Sanfilippo}  
\author[g]{Christian Schneider}
\author[c]{Nazario Tantalo}

\onbehalf{for the Extended Twisted Mass collaboration}

\affiliation[a]{Computation-based Science and Technology Research Center, The Cyprus Institute,\\20 Konstantinou Kavafi Street, 2121 Nicosia, Cyprus}

\affiliation[b]{Helmholtz-Institut Mainz, Johannes Gutenberg-Universität Mainz, 55099 Mainz, Germany \\ GSI Helmholtz Centre for Heavy Ion Research, 64291 Darmstadt, Germany}

\affiliation[c]{Dipartimento di Fisica and INFN, Universit\`a di Roma ``Tor Vergata",\\ 
Via della Ricerca Scientifica 1, I-00133 Rome, Italy}

\affiliation[d]{Dipartimento di Matematica e Fisica, Universit\`a Roma Tre and INFN, Sezione di Roma Tre,\\Via della Vasca Navale 84, I-00146 Rome, Italy}

\affiliation[e]{HISKP (Theory), Rheinische Friedrich-Wilhelms-Universit\"at Bonn,\\Nussallee 14-16, 53115 Bonn, Germany}

\affiliation[f]{Istituto Nazionale di Fisica Nucleare, Sezione di Roma Tre,\\Via della Vasca Navale 84, I-00146 Rome, Italy}

\affiliation[g]{Department of Physics, University of Cyprus, 20537 Nicosia, Cyprus}

\emailAdd{francesca.margari@roma2.infn.it}

\abstract{Low Mode Average (LMA) is a technique to improve the quality of the signal-to-noise ratio in the long time separation of Euclidean correlation functions. We report on its beneficial impact in computing the vector-vector light connected two-point correlation functions and
derived physical quantities in the mixed action lattice setup adopted by ETM collaboration. We focus on preliminary results of the computation within isospin symmetric QCD (isoQCD) of the $R$-ratio smeared with Gaussian kernels of widths down to $\sigma\sim250$~MeV, which is enough to appreciate the $\rho$ resonance around 770~MeV, using the Hansen-Lupo-Tantatlo (HLT) spectral-density reconstruction method.}

\FullConference{The 41st International Symposium on Lattice Field Theory (LATTICE2024)\\
 28 July - 3 August 2024\\
Liverpool, UK\\}

\begin{document}
\maketitle

\section{Introduction}{
The $R$-ratio between the $e^+e^-$ cross-section into hadrons with that into muons plays a significant role in elementary particle physics. Recently, its importance has increased due to its role in determining the leading hadronic contribution to the muon anomalous magnetic moment, $a_{\mu}$, by using the optical theorem and the dispersion relation. 

The $R$-ratio of electron-positron annihilation into hadrons, is defined as 
\begin{flalign}
R(\omega) = \frac{\sigma(e^+e^-\to \text{hadrons})}{\sigma(e^+e^-\to \mu^+\mu^-)} \; , 
\label{eq:R_omega}
\end{flalign}
$R(\omega)$ is a distribution and has to be probed by using smearing kernels. In Ref.~\cite{ExtendedTwistedMassCollaborationETMC:2022sta}, a first-principles lattice QCD investigation of the $R$-ratio smeared with Gaussian kernels is addressed, using the Hansen-Lupo-Tantalo (HLT) spectral-density reconstruction method~\cite{Hansen:2019idp}. 
Specifically, the authors computed the $R$-ratio convoluted with Gaussian smearing kernels, with widths ranging from 440 MeV to 630 MeV and center energies up to 2.5 GeV. 

The smeared $R$-ratio is extracted according to 
\begin{flalign}
R_\sigma(E) = \int_{0}^{\infty} d \omega \; G_\sigma (E-\omega) R(\omega) \; , 
\end{flalign}
with normalized Gaussian kernels, $G_\sigma(E-\omega)= \exp{(-(E-\omega)^2/2 \sigma^2)/\sqrt{2\pi\sigma^2}} $.
In Fig.~\ref{fig:R-ratio-paper}, the first-principles theoretical results of~\cite{ExtendedTwistedMassCollaborationETMC:2022sta} are compared with the corresponding quantity derived from the KNT19 compilation~\cite{Keshavarzi:2019abf} of $ R$-ratio experimental data smeared with the same Gaussian kernel. From a methodological standpoint, this result shows that it is feasible to study the $R$-ratio in Gaussian energy bins on the lattice with the level of accuracy necessary for precision tests of the Standard Model.

Here, we present a study of the smeared $R$-ratio at lattice spacings $a \simeq 0.057, 0.068, 0.080$ fm and spatial lattice sizes up to $L \simeq 5.5$ fm. We  reduce the Gaussian bin widths by increasing the statistical precision of our lattice correlators. To do this, we employ the HLT method combined with the Low Mode Averaging (LMA) technique. Indeed, the latter proves to be  highly beneficial for improving the signal-to-noise ratio in the long-time separations of Euclidean correlators.
Using this strategy, we present preliminary results for the smeared $R$-ratio with Gaussian kernels reduced to widths as small as $\sigma \sim 250$~MeV in the ETMC mixed action lattice setup \cite{Frezzotti:2003ni, Frezzotti:2004wz}. This is enough to appreciate the $\rho$ resonance around 770~MeV.
\begin{figure}[!h]
\centering
\includegraphics[width=0.68\textwidth]{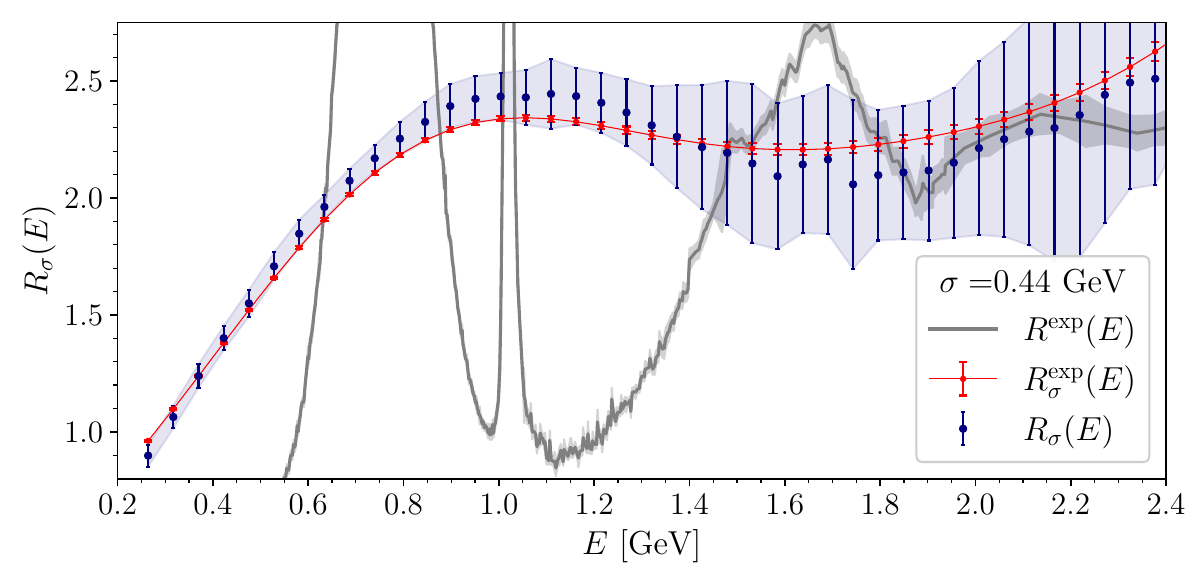}
\caption{The first-principle  lattice results of 2023 \cite{ExtendedTwistedMassCollaborationETMC:2022sta} compared to the smeared experimental $R$-ratio from KNT19 compilation \cite{Keshavarzi:2019abf}.
}\label{fig:R-ratio-paper}
\end{figure}
}

\section{Methodology}
We compute the two-point Euclidean correlator of the quark electromagnetic current given by
\begin{flalign}\label{eq:corr}
C(t)= - \frac{1}{3}\sum_{i=1}^3 \int \d^3x \mathrm{T} \bra{0} J_i(x) J_i(0) \ket{0} \; ,
\end{flalign}
where $J_\mu=\sum_{f}q_f \bar \psi_f \gamma_\mu \psi_f$ is the electromagnetic current, with $f=\{u,d,s,c,b,t\}$, $q_{u,c,t}=2/3$ and $q_{d,s,b}=-1/3$. We consider two lattice discretizations of the e.m. currents $J^{\mathrm{reg}}_{\mu, f}$ with reg = $\{$TM, OS$\}$ corresponding
to the so-called Twisted-Mass (TM) and Osterwalder-Seiler (OS) regularizations, see Ref.~\cite{ExtendedTwistedMassCollaborationETMC:2024xdf} for details.

This correlator is our primary observable directly computed via lattice simulations and it is connected to the $R$-ratio by the well known formula
\begin{flalign}
C(t)= \frac{1}{12 \pi^2} \int_{E_{\mathrm{th}}}^{\infty}\d \omega\, e^{-\omega t} \;\omega^2 R(\omega) \; .
\end{flalign}
where the threshold energy $E_{\mathrm{th}}$ is $2m_\pi$ in isoQCD.  

In the following, we first outline the LMA technique for computing the vector-vector light connected two-point correlation functions and then introduce the HLT spectral-density reconstruction method used in calculating the $R$-ratio smeared with Gaussian kernels.

\subsection{LMA method}{
LMA is a noise reduction technique in lattice QCD computations that enhances the signal-to-noise ratio of correlation functions at large distances, see \cite{Neff, Giusti:2004yp, DeGrand:2004qw, Borsanyi:2020mff} for details. 
The method isolates the contributions of the low eigenmodes of the Dirac operator from the remainder of the spectrum.

The eigenvectors are computed exactly and utilized to calculate all-to-all correlation functions, thereby extracting complete information from a gauge configuration. When a sufficient number of modes are included, the correlation function at large distances becomes dominated by the deflated contribution, significantly reducing statistical errors.

Here, we apply this approach to compute light-connected vector-vector two-point correlation functions within the framework of maximally twisted mass lattice regularization \cite{ExtendedTwistedMassCollaborationETMC:2024xdf}.
We introduce the Hermitian Dirac-Wilson matrix $Q_W$,
\begin{flalign}
Q_W \equiv \gamma_5 D_W = Q_W^\dagger \; , \qquad  Q_W v_j = \lambda_j v_j \; , \qquad 
\lambda_j {\rm~real} \; ,
\end{flalign}
where $D_W$ is the standard $\gamma_5$-Hermitian Wilson Dirac operator, and ($\lambda_j,v_j$) are eigenpairs of $Q_W$. Additionally, we define the twisted mass operator
\begin{flalign}
D_r = D_W + i r\mu \gamma_5 \; , \qquad Q_r \equiv \gamma_5 D_r = Q_W + i r \mu 
\;, \qquad \; Q_rv_j = (\lambda_j+ir\mu)v_j\;,
\label{eq:Q_r_def}
\end{flalign}
where $\mu$ is the bare positive twisted mass parameter and $r=\pm 1$ is the Wilson parameter sign in the twisted basis valence quark action.
The quark propagator $S_r$ satisfies the relation
\begin{flalign}
D_r(x,y) S_r(y,z)=\delta(x,z) \mathbb{1} \qquad \text{and}\qquad S_r = D_r^{-1} = Q_r^{-1} \gamma_5 \; , \label{eq_G_def}
\end{flalign}
where repeated indices are summed over, and color and spin indices are implied.

Quark propagators are often computed using stochastic methods, yielding
\begin{flalign}
S_{r,\eta}(x,y) = Q_r^{-1} \ket{\eta(x)} \bra{\eta(y)} \gamma_5 =  \ket{\phi_{r}^{\eta}(x)} \bra{ \eta(y)} \gamma_5 \; , 
\end{flalign}
where $\eta$ is properly defined stochastic source and $\phi_{r}^{\eta}$ is the solution to the linear system $Q_r\phi_{r}^{\eta}=\eta$. For bilinear two-point correlation functions we use as a stochastic source a time- and spin-diluted noise vector.

To introduce the deflated setup, we define the infrared (IR) projector, $P_{\rm IR}$, which projects onto the $N_v$ lowest modes of the Dirac operator, and the ultraviolet (UV), $P_{\rm UV}$, as
\begin{flalign}
P_{\rm IR}  =  \sum_{j=1}^{N_v} |v_j(x) \rangle \; \langle v_j(y)| 
\qquad\text{and} \qquad  P_{\rm UV} =   \mathbb{1} - P_{\rm IR}  \;.  
\label{eq:IR_UV_projectors}
\end{flalign}
The quark propagator can then be expressed as the sum of IR and UV contributions, where the IR component is computed exactly and the UV component stochastically, having
\begin{flalign}
\label{eq:propagator-split}
S_{r,\eta}^{\rm \,defl}(x,y) &= Q_r^{-1} P_{\rm IR} \gamma_5 + Q_r^{-1} P_{\rm UV}\ket{\eta(x)} \bra{\eta(y)}\gamma_5 \;\nonumber \\ &= \sum_{j=1}^{N_v} \frac{1}{\lambda_{j} + i r \mu} \ket{v_j(x)} \bra{v_j(y)} \gamma_5 +  \ket{\widetilde\phi_{r}^{\eta}(x)} \bra{ \eta(y)} \gamma_5 \; \\ &=
S_{r}^{\rm \,IR}(x,y) + S_{r,\eta}^{\rm \,UV}(x,y)\;,\nonumber
\end{flalign}
where $\widetilde\phi_{r}^{\eta}(x)$ is the solution to the linear system $Q_r\widetilde\phi_{r}^{\eta}=P_{\rm UV}\eta$. This decomposition ensures the efficient computation of the propagator by leveraging the exact treatment of the low modes and the stochastic evaluation of the high modes. 

We observe that the UV components of the purely stochastic propagator and the deflated propagator, $S_{r,\eta}^{\rm \,UV}(x,y)$, are identical when the same stochastic source is employed. Consequently, as anticipated, only the infrared part is improved, yielding
\begin{flalign}
\label{eq:propagator-split}
S_{r,\eta}^{\rm \,defl}(x,y)-S_{r,\eta}(x,y) &=  \sum_{j=1}^{N_v} \frac{1}{\lambda_{j} + i r \mu} \ket{v_j(x)} \bra{v_j(y)}\Big(\mathbb{1}- \ket{\eta(x)} \bra{\eta(y)}\Big) \gamma_5\;\nonumber\\&=S_{r}^{\rm \,IR}(x,y)-S_{r,\eta}^{\rm \,IR}(x,y).
\end{flalign}
Similarly, it can be shown that bilinear correlation functions computed using purely stochastic propagators and deflated propagators differ only in their purely infrared (IR) contributions, while the ultraviolet (UV) and mixed contributions remain identical. 
Spin dilution is essential to establish this result, as restricting the sources to a single spin component allows them to commute with the gamma matrices. For deflated two-point functions, the IR contributions are computed all-to-all, whereas in the non-deflated case, they are computed stochastically. Specifically,
\begin{flalign}
\label{eq:correlator-splitting}
C_{rr',\eta}^{\rm \,defl}(t)-C_{rr',\eta}(t) &= C_{rr'}^{\rm \,IR}(t)-C_{rr',\eta}^{\rm \,IR}(t)\;,
\end{flalign}
where $C_{rr',\eta}$, $C_{rr',\eta}^{\rm \,defl}$, $C_{rr',\eta}^{\rm \,IR}$, and $C_{rr'}^{\rm \,IR}$ are correlation functions computed using for propagators $S_{r,\eta}$, $S_{r,\eta}^{\rm \,defl}$, $S_{r,\eta}^{\rm \,IR}$, and $S_{r}^{\rm \,IR}$, respectively.

Using this relationship, we compute deflated two-point functions as
\begin{flalign}
\label{eq:correlator-split}
C_{rr',\eta}^{\rm \,defl}(t) &= C_{rr',\eta}(t)-C_{rr',\eta}^{\rm \,IR}(t)+C_{rr'}^{\rm \,IR}(t)\;,
\end{flalign}
where the standard stochastic correlation function and the IR part are computed separately, but both on the same sources. This approach is more efficient than explicitly constructing the deflated correlator containing IR, UV, and mixed terms, as it allows splitting the calculation into two distinct runs:
\begin{itemize}
    \item Stochastic Propagators: This run computes the purely stochastic correlation functions by inverting the Dirac operator using state-of-the-art techniques, such as multigrid solvers, on an optimal number of nodes.
    \item IR Contributions: This run computes the IR parts, which are more memory-intensive and require a larger number of nodes. However, this stage does not involve explicit inversion of the Dirac operator, relying instead on eigenvectors derived from eigensolvers.
\end{itemize}
The separation enables efficient scaling. Multigrid solvers, used in the first run, perform well on a smaller number of nodes, while the second run, dominated by eigenvector calculations and contractions, scales efficiently with additional nodes. A crucial requirement is that both runs produce identical stochastic sources, independent of the node configuration. To ensure this, we employ techniques similar to those in QUDA, assigning a unique seed to each lattice site.

\subsection{LMA performance}

We quantify the improvement in the signal-to-noise ratio achieved using the deflation technique by computing the following quantity:
\begin{flalign}
\label{eq:gain}
\mathrm{Gain} = \frac{\delta C^{\mathrm{no-defl}}(t)}{\delta C^{\mathrm{defl}}(t)} \times \sqrt{\frac{N_U^{\mathrm{no-defl}}}{N_U^{\mathrm{defl}}}} \times \sqrt{\frac{N_\eta^{\mathrm{no-defl}}}{N_\eta^{\mathrm{defl}}}}  \; ,
\end{flalign}
where $C^{\mathrm{no-defl}}$ denotes the vector-vector light connected two-point correlator taken from Ref.~\cite{windows}. The terms $N_U$ and $N_\eta$ represent the numbers of gauge configurations and stochastic sources used in each setup, respectively.

To determine the optimal gain, we analyze the results using various numbers of eigenvectors and stochastic sources. In Fig.~\ref{fig:gain_nvec}, we show the gain, as defined in Eq.~\eqref{eq:gain}, computed for the deflated two-point functions in Eq.~\eqref{eq:correlator-split} on a $64^3 \times 128$ lattice (with $a \simeq 0.080$ fm). The study considers $N_{\rm eig} = 200, 300, 400, 500$, using $N_U = 56$ gauge configurations and $N_\eta = 256$ time-wall stochastic sources. The optimal choice is $N_{\rm eig}=400$, as the average gain in this case is statistically compatible with that obtained using 500 eigenvectors, while requiring fewer computational resources.
A similar study was performed to determine the optimal number of stochastic sources. We find that the stochastic noise saturates with approximately 1000 time-wall stochastic sources per configuration.

For the optimal choices above of $N_{\rm eig}$ and $N_\eta$, we present in Fig.~\ref{fig:corr_gain_plot} a comparison of the correlators obtained with and without the LMA technique for both TM and OS lattice regularizations, using a larger statistics: $N_U = 790$
and $N_\eta=1024$ without LMA (as used in~\cite{windows}) and 
a smaller statistics with the more efficient LMA-based method.
On the left, we report the relative gain in the signal-to-noise ratio: we observe a reduction of
the error by a factor of 3.63(15) and 3.57(16) for TM and OS, respectively. On the other hand, the computational cost has increased by a factor of about 2.7(2). Thus, for TM the net gain in computer time to reach a given accuracy is a factor of 5(1).

To achieve the same signal-to-noise ratio gain for the other two lattice spacings, $a \simeq 0.057, 0.068$ fm, we analyze the frequency histogram $(\Delta N/\Delta \tilde\lambda) (\tilde\lambda) $ of eigenvalues $\tilde\lambda$ of $(Q_W^2 + \mu^2)^{1/2}$ shown in Fig.~\ref{fig:histogram}. Given that, based on the chiral properties of QCD \cite{BANKS1980103, Leutwyler:1992yt, Luscher:2007se, Giusti:2004yp} 
in the large volume
limit $(\Delta N/\Delta \tilde\lambda) (0) \propto \Lambda_{\rm QCD}^3 L^3 T $, we assume the number $K$
of eigenvalues $\tilde\lambda$ below a certain threshold $\tilde\lambda_{\rm thrs}$, which for practical convenience is set about $8 \mu_\ell^{\rm sea}$, to scale proportionally to 
the spacetime volume $L^3 T$ in physical units, independently (to a first approximation) of $a$. 

Table~\ref{tab:ensembles} provides a comprehensive summary of the ETMC gauge ensembles used in this study, including all relevant LMA setup parameters.

\begin{figure}[!t]
\centering
\includegraphics[width=0.6\textwidth]{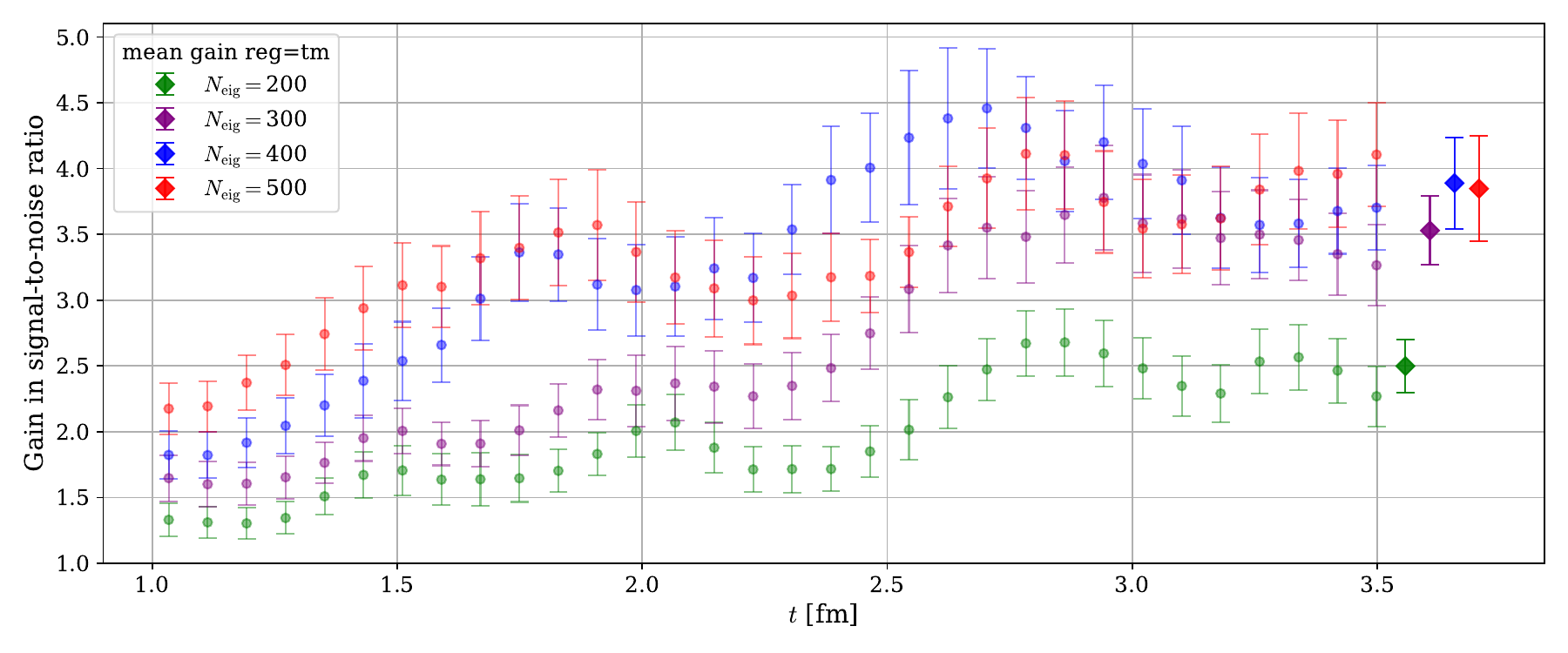}
\caption{\label{fig:gain_nvec} The gain in the signal-to-noise ratio defined in Eq.~\eqref{eq:gain}, computed for the deflated two-point correlators on a $64^3 \times 128$ lattice with $L = 5.1$ fm, for TM lattice regularization. Results are shown for $N_{\rm eig} = 200, 300, 400, 500$, $N_U = 56$ and $N_\eta = 256$. The optimal choice is $N_{\rm eig}=400$, achieving a gain statistically compatible with that for 500 eigenvectors while requiring fewer computational resources.}
\end{figure}

\begin{figure}[!h]
\centering
\includegraphics[width=0.9\textwidth]{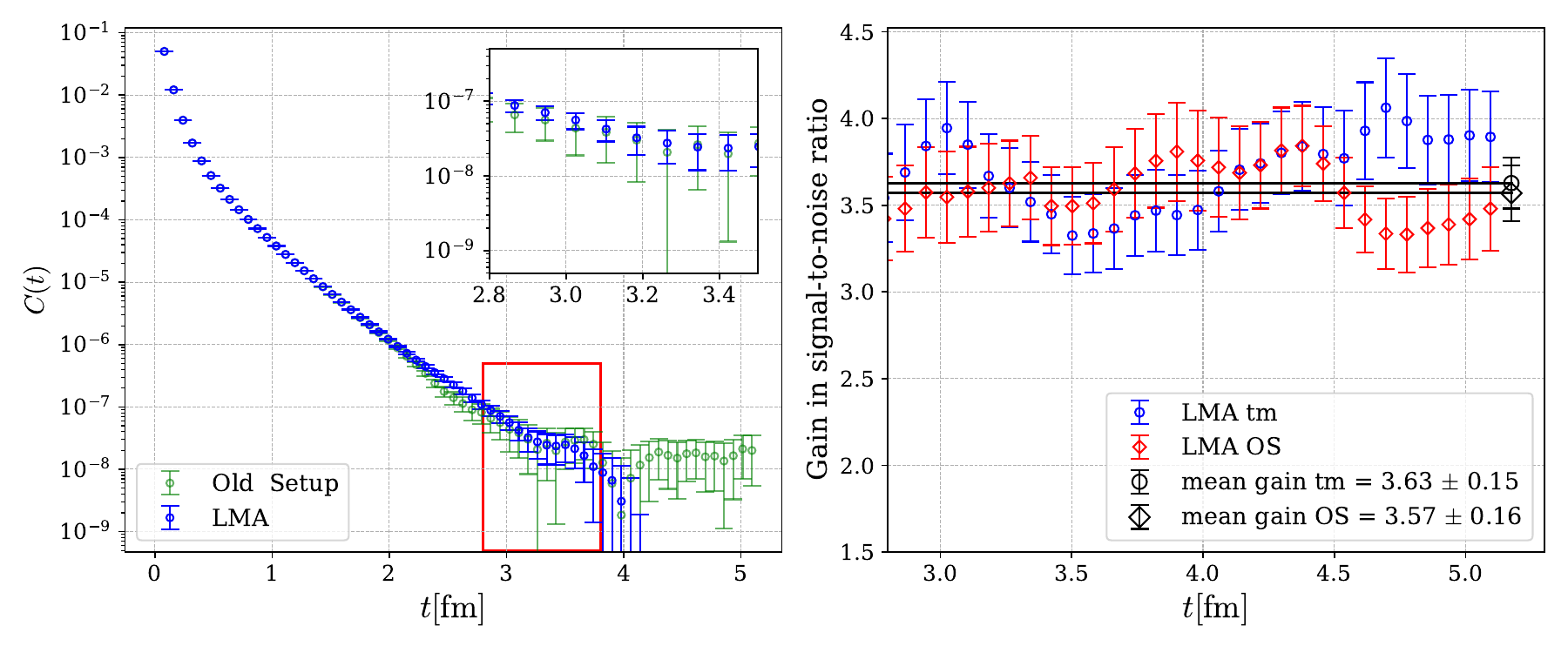}
\caption{\label{fig:corr_gain_plot} Left: Comparison of the two-point e.m. current correlator in isoQCD, on a $64^3 \times 128$ lattice with $L = 5.1$ fm, by using the old setup, with no LMA and $(N_U,N_\eta)=(780,1000)$ and the new setup, with LMA and a reduced statistics 
$(N_U,N_\eta)=(158,1000)$ - the full statistics data are blinded, further details will be given in Ref.~\cite{ETMC_light}. 
Right: the gain in the signal-to-noise ratio defined in Eq.~\eqref{eq:gain}. We reach a reduction of the error by a factor of 3.63(15) and 3.57(16) for TM and OS, respectively.}
\end{figure}

\begin{figure}[!t]
\centering
\includegraphics[width=0.7\textwidth]{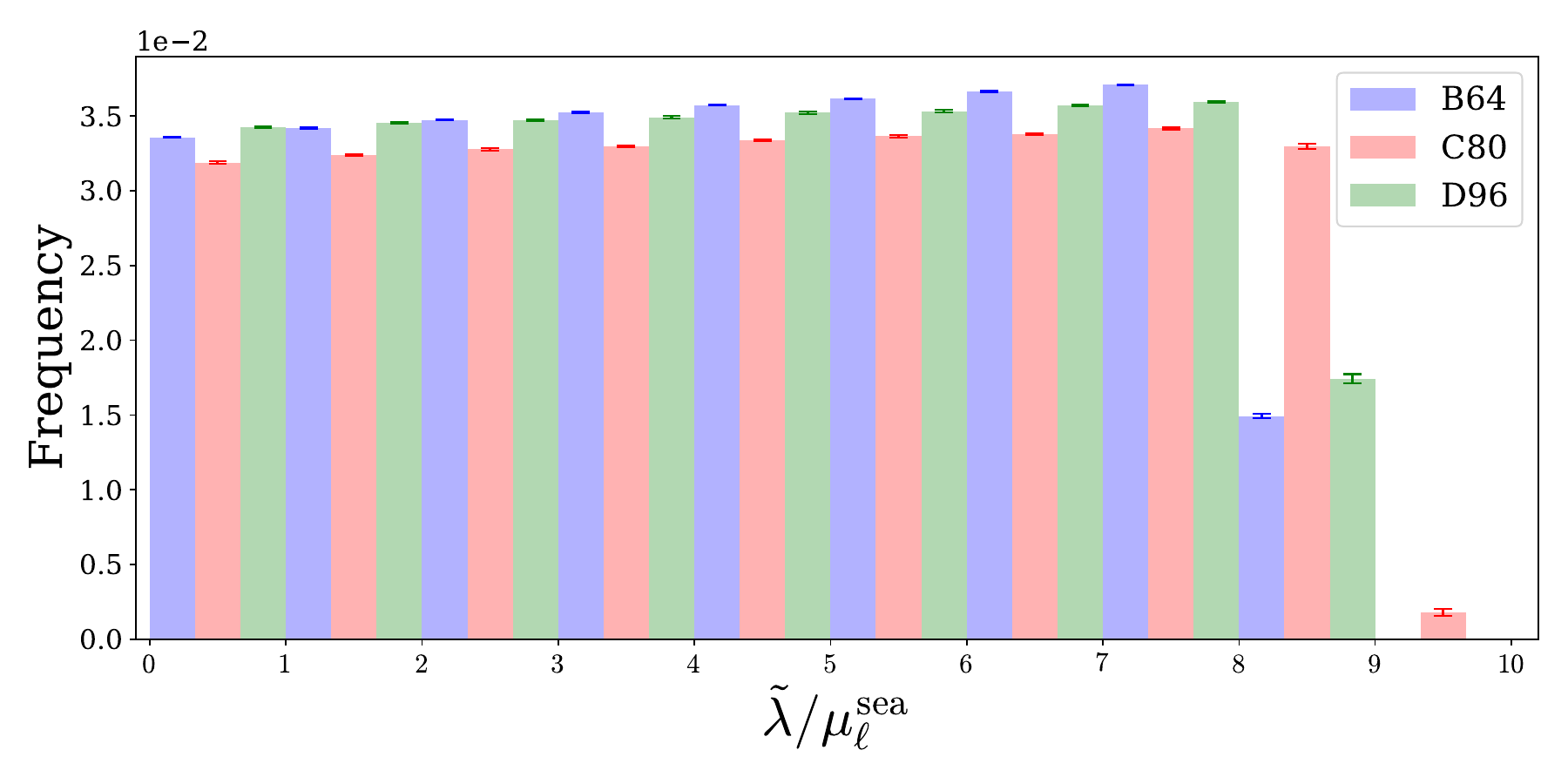}
\caption{The frequency histogram $(\Delta N/\Delta \tilde\lambda) (\tilde\lambda) $ of eigenvalues $\tilde\lambda$ of $(Q_W^2 + \mu^2)^{1/2}$ for B64, C80 and D96 ensembles (see Tab.~\ref{tab:ensembles} for details on the gauge ensembles). 
}
\label{fig:histogram}
\end{figure}
}

\begin{table}[!t]
\begin{center}
\begin{tabular}{cccccc}
Ensemble name (short) & $(L/a)^3 \times T/a$ & $a$[fm] & $N_U$ & $N_{\rm eig}$ & $N_\eta$ \\
\hline
cB211.072.64 (B64) & $64^3 \times 128$ & 0.07951(4) & 790 & 400 & 1024 \\
cC211.060.80 (C80) & $80^3 \times 160$ & 0.06816(8) & 550 & 530 & 960 \\
cD211.054.96 (D96) & $96^3 \times 192$ & 0.05688(6) & 400 & 530 & 960 \\ 
\end{tabular}
\caption{\label{tab:ensembles}%
ETMC ensembles \cite{poster-Bartosz} used for two-point vector correlators in isoQCD.}
\end{center}
\end{table}

\subsection{Basics of the HLT approach}
The determination of $R_\sigma(E)$ on the lattice, with controlled statistical and systematic uncertainties, is achievable using the HLT method outlined in Ref.~\cite{Hansen:2019idp}. According to this approach, the smearing kernels are approximated as 
\begin{flalign}
K(\omega; \mathbf{g}) = \sum_{\tau=1}^{\tau_{\rm max}} g_{\tau} e^{-a \omega \tau} \; ,
\end{flalign}
where $\tau$ is an integer variable and $a$ is the lattice spacing. The distance between the target kernel and its representations in terms of the coefficients $g_{\tau}$ is measured by the functionals
\begin{flalign}
A_{\rm n} [\mathbf{g}] = \int_{E_0}^{\infty} \d \omega  \ w_{\mathrm{n}}(\omega) \left| K(\omega; \mathbf{g}) - \frac{12\pi^2G_\sigma(E-\omega)}{\omega^2}  \right|^2 \, , 
\end{flalign}
that, for weight-functions $w_{n}>0$, correspond to a class of weighted $L_2$-norms in functional space. The \textbf{g} coefficients result from minimizing
\begin{flalign}
W_{\rm n} [\lambda, \mathbf{g}] = \frac{A_{\rm n}[\mathbf{g}]}{A_{\rm n}[\mathbf{0}]} + \lambda   B [\mathbf{g}], \quad \text{where} \ B [\mathbf{g}] = \sum_{\tau_{1,2}=1}^{T/a} g_{\tau_1}g_{\tau_2} \mathrm{Cov}_{\tau_1 \tau_2} \; ,
\end{flalign}
where $\lambda$ is a parameter varied/optimized in  stability analysis. For a complete and detailed description, we refer the reader to the Supplementary Material section in Ref.~\cite{ExtendedTwistedMassCollaborationETMC:2022sta}.

\section{Results for smeared $R$-ratio using LMA}{
In Fig.~\ref{fig:r-ratios}, we present preliminary blinded results for the connected contribution of the $u/d$ quark mass to the smeared $R_{\sigma}(E)$, evaluated for central energies in the range $E \in [0.2, 1.6]$ GeV with a fixed resolution of $\sigma = 250$ MeV. Here, the resolution $\sigma$ refers to the standard deviation of the Gaussian kernel applied to smear $R(E)$.

Specifically, in Fig.~\ref{fig:r-ratios} (left), we show the results for $R^{\ell \ell}_{\sigma}(E)$ obtained through a spectral reconstruction analysis of correlators evaluated on the B64 ensemble with 780 configurations, both in the new LMA setup and the old setup. By increasing the statistical precision of the vector-vector light connected two-point correlators with the LMA technique, we reduce the relative error on $R^{\ell \ell}_{\sigma}(E)$ from $11\%$ to $3\%$ at an energy of approximately 770 MeV and a resolution of $\sigma = 250$ MeV. This improvement enables us to resolve the $\rho$ resonance and facilitates a direct phenomenological comparison with experimental data.

Additionally, in Fig.~\ref{fig:r-ratios} (right), we present results for $R^{\ell \ell}_{\sigma}(E)$ obtained by combining the HLT + LMA technique for the B64 and C80 ensembles, considering both TM and OS lattice regularizations. 
These results, derived from independent simulations, are compatible with each other and, with the achieved precision, we are able to distinguish lattice artifacts at low energies that are inherent to the two lattice regularizations we employed.

\begin{figure} 
\centering \includegraphics[width=7.5cm,height=5.5cm]{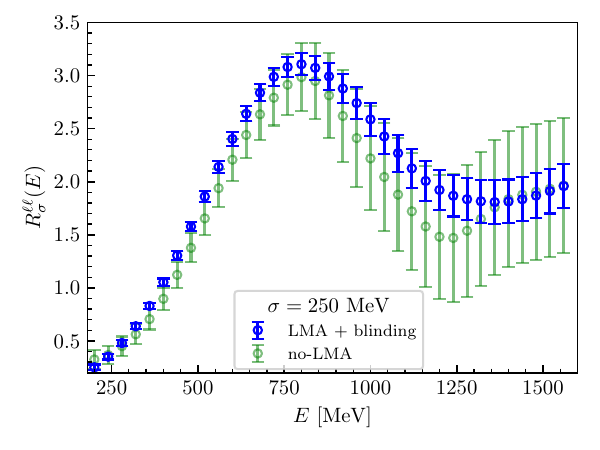} \includegraphics[width=7.5cm,height=5.5cm]{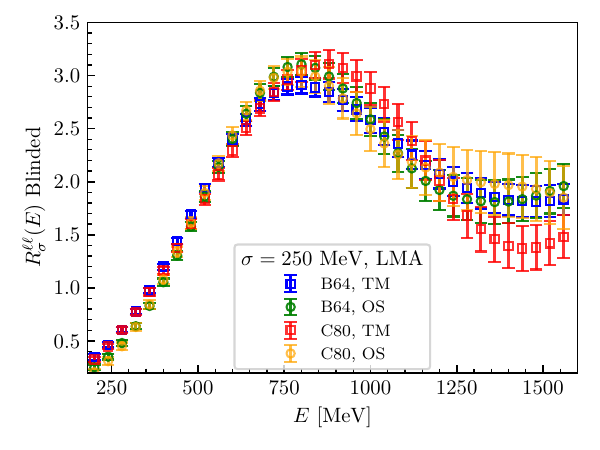} 
\caption{\label{fig:r-ratios} Preliminary blinded values of the connected contribution of the $u/d$ quark mass to the smeared $R_{\sigma}(E)$ with central energies up to 1.6 GeV and a fixed resolution of $\sigma = 250$ MeV. Left: Results for the B64 ensemble, obtained through a spectral reconstruction analysis using the HLT method as described in~\cite{Hansen:2019idp}, applied to correlators computed in both the new LMA setup and the old setup. Right: Results obtained using the combined HLT + LMA technique for the B64 and C80 ensembles, comparing TM and OS lattice regularizations.} 
\end{figure}
}

\section*{Acknowledgments}
We thank all members of ETMC for the most enjoyable collaboration. 

 S.B.~and F.P.~received financial support from the Inno4scale project, which received funding from the European High-Performance Computing Joint Undertaking (JU) under Grant Agreement No.~101118139. The JU receives support from the European Union's Horizon Europe Programme.
A.D.S, A.E., R.F., G.G., F.M., F.S.~and N.T. are supported by the Italian Ministry of University and Research (MUR) under the grant PNRR-M4C2-I1.1-PRIN 2022-PE2 Non-perturbative aspects of fundamental interactions, in the Standard Model and beyond F53D23001480006 funded by E.U.-NextGenerationEU. M.G. is supported by the Deutsche Forschungsgemeinschaft (DFG,
German Research Foundation) as part of the CRC 1639 NuMeriQS – project no. 511713970. 
F.S. is supported by ICSC – Centro Nazionale di Ricerca in High Performance Computing, Big Data and Quantum Computing, funded by European Union -NextGenerationEU and by Italian  Ministry of University and Research (MUR) projects FIS\_00001556 and PRIN\_2022N4W8WR.
C.S. has received financial support under the AQTIVATE EJD from the European Union’s research and innovation programme under the Marie Skłodowska-Curie Doctoral Networks action and Grant Agreement No 101072344.

The open-source packages tmLQCD~\cite{Jansen:2009xp,Abdel-Rehim:2013wba,Deuzeman:2013xaa,Kostrzewa:2022hsv} LEMON~\cite{Deuzeman:2011wz}, DD-$\alpha$AMG~\cite{Frommer:2013fsa,Alexandrou:2016izb,Bacchio:2017pcp,Alexandrou:2018wiv}, QPhiX~\cite{joo2016optimizing,Schrock:2015gik} and QUDA~\cite{Clark:2009wm,Babich:2011np,Clark:2016rdz} have been used in the ensemble generation.

The authors gratefully acknowledge the Gauss Centre for Supercomputing e.V.~(www.gauss-centre.eu) for funding this project by providing computing time on the GCS Supercomputers SuperMUC-NG at Leibniz Supercomputing Centre and JUWELS~\cite{JUWELS, JUWELS-BOOSTER} at Juelich Supercomputing Centre.
The authors acknowledge the Texas Advanced Computing Center (TACC) at The University of Texas at Austin for providing HPC resources (Project ID PHY21001).
The authors gratefully acknowledge PRACE for awarding access to HAWK at HLRS within the project with Id Acid 4886.
We acknowledge the Swiss National Supercomputing Centre (CSCS) and the EuroHPC Joint Undertaking for awarding this project access to the LUMI supercomputer, owned by the EuroHPC Joint Undertaking, hosted by CSC (Finland) and the LUMI consortium through the Chronos programme under project IDs CH17-CSCS-CYP and CH21-CSCS-UNIBE as well as the EuroHPC Regular Access Mode under project ID EHPC-REG-2021R0095.
We are grateful to CINECA and EuroHPC JU
for awarding this project access to Leonardo 
supercomputing facilities hosted at CINECA. 
We gratefully acknowledge EuroHPC JU for the computer
time on Leonardo-Booster provided to us through the 
Extreme Scale Access Call grant EHPC-EXT-2024E01-027.
We gratefully acknowledge
CINECA for the provision of GPU time under the specific initiative INFN-LQCD123 and IscrB S-EPIC.

\bibliographystyle{JHEP}
\bibliography{biblio}

\end{document}